\documentclass{article}

\usepackage[bindingoffset=0cm,a4paper,centering,textheight=200mm,textwidth=120mm]{geometry}
  
\usepackage{amssymb}
\usepackage{amsmath}
\usepackage{amsthm}
\usepackage{stmaryrd}
\usepackage[usenames,dvipsnames]{xcolor}
\usepackage{proof-dashed}
\usepackage{tikz}
\usetikzlibrary{arrows}
\usetikzlibrary{positioning}
\usepackage{import}
\usepackage{wrapfig}
\usepackage[all]{xy}
\usepackage{url}

\definecolor{mygrey}{rgb}{.9,.9,.9}

\newtheorem{theorem}{Theorem}

\newtheorem{example}{Example}

\newcommand{\m}[1]{\ensuremath{\mathsf{#1}}}
\newcommand{\pid}[1]{\m{#1}}
\newcommand{\pids}[1]{\til {\pid #1}}

\newcommand{\code}[1]{\texttt{\upshape#1}}
\newcommand{\nil}{\boldsymbol 0}

\newcommand{\com}[2]{#1\;\code{-\hspace{-0.3mm}>}\;#2}
\newcommand{\intro}[3]{#1.{#3}\;\code{-\hspace{-0.3mm}>}\;#2}
\newcommand{\tell}[3]{#1:{#3}\;\code{<\hspace{-0.3mm}-\hspace{-0.3mm}>}\;#2}
\newcommand{\gencom}{\com{\pid p.e}{\pid q}}
\newcommand{\sel}[3]{\com{#1}{#2 [#3]}}
\newcommand{\gensel}{\sel{\pid p}{\pid q}{l}}

\newcommand{\cond}[3]{\m{if}\, #1 \, \m{then} \, #2 \, \m{else} \, #3}
\newcommand{\eqcom}[2]{#1 \! \stackrel{\code{<\!-}}{\code{=}}\! #2}

\newcommand{\gencond}{\cond{\eqcom{\pid p}{\pid q}}{C_1}{C_2}}
\newcommand{\genintro}{\intro{\pid p}{\pid q}{\pid r}}
\newcommand{\gentell}{\tell{\pid p}{\pid q}{\pid r}}
\newcommand{\start}[3][]{#2 \, \m{start}^{#1} \, #3}

\newcommand{\genstart}{\pid p\, \m{starts}\, \pid q}

\newcommand{\rec}[3]{\m{def} \, #1  =  #2 \, \m{in} \, #3}
\newcommand{\genrec}{\rec{X}{C_2}{C_1}}
\newcommand{\genrecp}{\rec{X(\pids p)}{C_2}{C_1}}
\newcommand{\pn}{\m{pn}}

\newcommand{\callp}[2]{#1 \langle #2 \rangle}
\newcommand{\gencall}{X}
\newcommand{\gencallp}{\callp X{\pids p}}
\definecolor{light-gray}{gray}{0.928}

\newcommand{\pcont}{\ast}
\newcommand{\eval}[2]{{#1}\downarrow{#2}}
\newcommand{\til}{\tilde}

\newcommand{\amend}{\m{Amend}}
\newcommand{\knows}[3][G]{\ensuremath{{#2}\stackrel{#1}{\longleftrightarrow}{#3}}}
\newcommand{\update}[3][G]{\ensuremath{{#1}\cup\{{#2}\leftrightarrow{#3}\}}}
\newcommand{\know}[3][G]{\ensuremath{{#2}\stackrel{#1}{\rightarrow}{#3}}}
\newcommand{\updates}[3][G]{\ensuremath{{#1}\cup\{{#2}\rightarrow{#3}\}}}

\newcommand{\NN}{\mathbb N}

\newcommand{\altenc}[3]{\{\!\!\{#1\}\!\!\}^{#2}_{#3}}

\newcommand{\rname}[2]{\ensuremath{\left\lfloor\mbox{{#1}$|${#2}}\right\rceil}}

\newcommand{\precongr}{\preceq}

\newcommand{\smallpar}[1]{\smallskip

\noindent \textbf{\emph{#1}}}

\newcommand{\acspar}{\hspace{0.8mm} | \hspace{0.8mm} }

\newcommand{\delsel}[1]{(\![{#1}]\!)}

\newcommand{\async}[2]{\altenc{#1}{}{#2}}

\begin{document}

\title{That's Enough: Asynchrony with Standard Choreography Primitives}
\author{Lu\'\i s Cruz-Filipe and Fabrizio Montesi\\
  Dept.\ Mathematics and Computer Science\\
  University of Southern Denmark\\
  \texttt{\{lcf,fmontesi\}@imada.sdu.dk}}
\date{}

\maketitle

\begin{abstract}

Choreographies are widely used for the specification of concurrent and distributed software architectures.
Since asynchronous communications are ubiquitous in real-world systems, previous works have proposed different 
approaches for the formal modelling of asynchrony in choreographies.
Such approaches typically rely on ad-hoc syntactic terms or semantics for capturing the concept of messages in transit, 
yielding different formalisms that have to be studied separately.

In this work, we take a different approach, and show that such extensions are not needed to reason about asynchronous 
communications in choreographies. Rather, we demonstrate how a standard choreography calculus already has all the 
needed expressive power to encode messages in transit (and thus asynchronous communications) through the primitives of 
process spawning and name mobility.
The practical consequence of our results is that we can reason about real-world systems within a
choreography formalism that is simpler than those hitherto proposed.


\end{abstract}

\section{Introduction}
\label{sec:intro}

Today, concurrent and distributed systems are widespread. Multi-core hardware and large-scale networks represent the 
norm rather than the exception.
However, programming such systems is challenging, because it is difficult to program correctly the intended 
interactions among components executed concurrently (e.g., services).
Empirical investigations of bugs in concurrent and distributed software~\cite{LLLG16,LPSZ08} reveal that most errors 
are due to: deadlocks (e.g., a component that was supposed to be ready for interaction at a given time is actually 
not); violations of atomicity intentions (e.g., a component is performing some action when not intended to); or, 
violations of ordering intentions (some components perform the right actions, but not when intended).
If the design and implementation of a concurrent system are initially difficult, they get 
even harder as the system evolves and has to be maintained.
Without proper tool support, introducing new actions at components may have 
unexpected effects due to side-effects.

To mitigate this problem, \emph{choreographies} can be used as high-level formal specifications of the 
intended interactions among components~\cite{BB11,BPMN,BGGLZ06,CHY12,HYC16,LGMZ08,QZCY07,WSCDL}.
\begin{example}
\label{ex:abs}
We use a choreography to define a scenario where a buyer, Alice ($\pid a$), purchases a product from a seller ($\pid s$) 
through her bank ($\pid b$).
\begin{align*}
1.\quad & \com{\pid a.\mathit{title}}{\pid s};\\
2.\quad & \com{\pid s.\mathit{price}}{\pid a};\\
3.\quad & \com{\pid s.\mathit{price}}{\pid b};\\
4.\quad & \m{if} \ \eqcom{\pid b}{\pid a} \ \m{then}\\
5.\quad & \qquad \sel{\pid b}{\pid s}{\mathit{ok}};\ \sel{\pid b}{\pid a}{\mathit{ok}};\\
6.\quad & \qquad \com{\pid s.book}{\pid a};\\
7.\quad & \m{else} \ \, \sel{\pid b}{\pid s}{\mathit{ko}}; \ \sel{\pid b}{\pid a}{\mathit{ko}}
\end{align*}
In Line 1, the term $\com{\pid a.\mathit{title}}{\pid s}$ denotes an interaction whereby $\pid a$ communicates the title of the 
book that Alice wishes to buy to $\pid s$.
The seller then sends the price of the book to both $\pid a$ and $\pid b$. In Line 4, $\pid a$ sends the price she 
expects to pay to $\pid b$, which confirms that it is the same amount requested by $\pid s$ (stored internally at $\pid 
b$).
If so, $\pid b$ notifies both $\pid s$ and $\pid a$ of the successful transaction (Line 5) and $\pid s$ sends the book 
to $\pid a$ (Line 6). Otherwise, $\pid b$ notifies $\pid s$ and $\pid a$ of the failure (Line 7) and the choreography 
terminates.
\end{example}
%
Choreographies are the foundations of an emerging development paradigm, called Choreographic 
Programming~\cite{M13:phd,M15},
where an automatic projection procedure is used to synthesise a 
set of compliant local implementations (the implementations of the 
single components) from a choreography~\cite{CHY12,LGMZ08,QZCY07}.
This procedure is formally proven to be correct, preventing deadlocks, ordering errors, and atomicity 
violations. This ensures, critically, that updates to either the choreography or the local implementations do not 
introduce bugs and that developers always know what communications their systems will enact (by looking at the 
choreography).
In the previous example, the implementation inferred for, e.g., Alice ($\pid a$), would be: send the book title to 
$\pid s$; receive the price from $\pid s$; send the price to $\pid b$ for confirmation; await the success/failure 
notification from $\pid b$; in case of success, receive the book from $\pid s$.

Choreography languages come in all sizes and flavours, with different sets of primitives inspired by practical 
applications, such as adaptation~\cite{PGGLM15,PGLMG14}, channel mobility~\cite{CM13,chor:website}, or web 
services~\cite{BPMN,CHY12,WSCDL}.
However, this multiplicity makes it increasingly difficult to reuse available theory and tools, because of the 
differences and redundancies among these models.
For this reason, we previously introduced the model of Core
Choreographies (CC)~\cite{ourstuff}, a minimal and representative theoretical model of Choreographic Programming.
In CC, components are modelled as concurrent processes that run independently and possess own memory, inspired by 
process calculi~\cite{SW01}. Example~\ref{ex:abs} is written in the syntax of CC described in \S~\ref{sec:mc}.

In this paper, we are interested in studying asynchronous communications in choreographies.
As a motivation, consider the two communications in Lines 2 and 3 of Example~\ref{ex:abs}: typically, in a 
realistic system, we would expect $\pid s$ to send the price to $\pid a$ and then immediately proceed to sending it also 
to $\pid b$, without waiting for $\pid a$ to receive its message.
Typically, asynchronous communications are formalised in choreography models by defining ad-hoc extensions 
to their syntax and semantics~\cite{CM13,DY13,HYC16,LGMZ08,MY13,MYH09}, causing a substantial amount of duplication in 
their technical developments (many of which are even incompatible with each other).
%
%

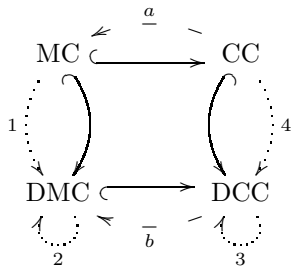
\begin{figure}
\[
\xymatrix@C+2ex@R+3ex{
  \mbox{\ MC\ }\raisebox{-1mm}{\rule{0mm}{0mm}}
  \ar@{^{(}->}[]+R-<0px,3px>;[r]+L-<0px,3px>
  \ar@{_{(}->}@/^1pc/[d]
  \ar@{.>}@/_1pc/[d]_{1}
  &
  \mbox{\ CC\ }\raisebox{-1mm}{\rule{0mm}{0mm}}
  \ar@{^{(}->}@/_1pc/[d]
  \ar@{.>}@/^1pc/[d]^{4}
  \ar@{-->}@/_1pc/[l]_{a}
  \\
  \mbox{\ DMC\ }\raisebox{-1mm}{\rule{0mm}{0mm}}
  \ar@{_{(}->}[]+R+<0px,3px>;[r]+L+<0px,3px>
  \ar@{.>}@(dr,dl)[]^{2}
  &
  \mbox{\ DCC\ }\raisebox{-1mm}{\rule{0mm}{0mm}}
  \ar@{.>}@(dr,dl)[]^{3}
  \ar@{-->}@/^1pc/[l]^{b}
}
\]
\caption{Choreography calculi and encodings.}
\label{fig:ouroboros}
\end{figure}

Unfortunately, there are still no foundational studies that provide an elegant and general understanding of asynchrony 
in choreographies. Here, we pursue such a study in the context of CC. We depict our overall development in 
Figure~\ref{fig:ouroboros}, and describe it in the following.

%

We first present our development for the 
computational fragment of CC, called Minimal Choreographies (MC)~\cite{ourstuff}.
We take inspiration from how asynchrony is modelled in foundational process models, specifically the 
$\pi$-calculus~\cite{MPW92}. The key idea there is to use processes to represent messages in transit, allowing the 
sender to proceed immediately after having sent a message without having to synchronise with the receiver~\cite{SW01}.
In an asynchronous system, there is no bound to the number of messages that could be transiting in the network;
this means that MC is not powerful enough for our purposes, because it can only capture a finite number of 
processes (the same holds for CC).
For this reason, we extend MC with two standard notions, borrowed from process calculi and previous choreography 
models: process spawning -- the ability to create new processes at runtime -- and name mobility -- the ability to send 
process references, or names. We call this new language Dynamic Minimal Choreographies (DMC). MC is a strict 
sub-language of DMC, denoted by the arrow $\hookrightarrow$ on the left-hand side of Figure~\ref{fig:ouroboros}. 
In general, all arrows of shape $\hookrightarrow$ in that figure denote (strict) language inclusion.

The dotted arrow ($1$) in Figure~\ref{fig:ouroboros} is the cornerstone of our development: every choreography in MC 
can be encoded in an asynchronous implementation in DMC, by using auxiliary processes to represent messages in 
transit. Since DMC extends MC with new primitives, it makes sense to extend this encoding to the whole language of DMC 
($2$).
This syntactic interpretation of asynchrony in choreographies is our main contribution. Specifically, our results show 
that asynchronous communications can be modelled in choreographies using well-known notions, i.e., process spawning and 
name mobility (studied, e.g., in~\cite{CM13,ourPCstuff}), without the need for ad-hoc constructions.
Coming back to the title: we already have enough.

The fact that our encoding can be extended from MC to DMC is evidence that our approach is robust, and the simplicity 
of DMC makes it a convenient foundational calculus to use in future developments of choreographies.
However, one of the expected advantages of using a foundational theory such as DMC for capturing asynchrony is indeed 
that we can reuse existing formal techniques based on standard primitives for choreographies. (This is a common 
scenario in $\pi$-calculus, where many techniques apply to its sub-languages~\cite{SW01}.)
We show an example of such reuses.
Core Choreographies (CC)~\cite{ourstuff} is MC with the addition of a primitive for communicating choices explicitly as 
messages, called selection~\cite{CHY12,HVK98,HYC16,YHNN13} (the terms in Lines 5 and 7 in Example~\ref{ex:abs} are 
selections).
An important property of CC is that selections can be encoded in the simpler language MC -- the dashed arrow ($a$) in 
Figure~\ref{fig:ouroboros}.
What happens if we add selections to DMC?
Ideally, the resulting calculus (called Dynamic Core Choreographies, or DCC) should \emph{both} have an asynchronous 
interpretation through the techniques introduced in this paper and still possess the property that selections 
are encodable using the simpler language DMC.
This is indeed the case. We extend our encoding to yield an interpretation of asynchronous selections, yielding ($3$) 
and ($4$). The second property (encodability of selections in DCC) follows immediately from language inclusion, giving
us ($b$) for free.


\section{Background}
\label{sec:mc}

We briefly introduce CC and MC, from~\cite{ourstuff}, and summarise their key properties.

The syntax of CC is given in Figure~\ref{fig:cc_syntax}, where $C$ ranges over choreographies.
\begin{figure}
  \begin{align*}
    C & ::= \nil \acspar \eta; C \acspar \gencond \acspar  \genrec \acspar \gencall
    \\
    \eta & ::= \gencom \acspar \gensel
    \hspace{15.4mm}
    e ::= v \acspar \pcont \acspar \ldots
\end{align*}
\caption{Core Choreographies, Syntax.}
\label{fig:cc_syntax}
\end{figure}
Processes ($\pid p,\pid q,\ldots$) run in parallel, and each process stores a
value in a local memory cell.\footnote{In the original presentation, values were restricted to natural numbers; we drop 
this restriction here since it is orthogonal to our development.}
Each process can access its own value using the syntactic construct $\pcont$, but it cannot read the contents of another process.
Term $\eta;C$ is an interaction between two processes, read
``the system may execute $\eta$ and proceed as $C$''.
In a value communication $\gencom$, $\pid p$ sends its local evaluation of expression $e$ to $\pid q$, which stores the
received value.
In a label selection $\gensel$, $\pid p$ communicates label $l$ to $\pid q$.
The set of labels is immaterial, as long as it contains at least two elements.
In a conditional $\gencond$, $\pid q$ sends its value to $\pid p$, which checks if the received value is equal to its own; the choreography proceeds as $C_1$, if that is the case, or as $C_2$, otherwise.
In all these actions, the two interacting processes must be different.
Definitions and invocations of recursive procedures ($X$) are standard.
The term $\nil$ is the terminated choreography.

The semantics of CC uses reductions of the form $C,\sigma \to C',\sigma'$, where the total state function $\sigma$ maps each process name to its value.
We use $v$, $w$, $\ldots$ to range over values.
The reduction relation $\to$ is defined by the rules given in Figure~\ref{fig:cc_semantics}.

\begin{figure}
\begin{eqnarray*}
&\infer[\rname{C}{Com}]
{
	\gencom;C,\sigma
	\to
	C, \sigma[\pid q \mapsto v]
}
{
	\eval{e[\sigma(\pid p)/\pcont]}v
}
\\[1ex]
&
\infer[\rname{C}{Sel}]
{
	\gensel;C, \sigma \to C, \sigma
}
{}
\\[1ex]
&\infer[\rname{C}{Cond}]
{
	\gencond, \sigma  \to    C_i, \sigma
}
{
	i = 1 \ \text{if } \sigma(\pid p) = \sigma(\pid q),\ 
	i = 2 \ \text{o.w.}
}
\\[1ex]
&\infer[\rname{C}{Ctx}]
{
	\genrec, \sigma \ \to \ 
	\rec{X}{C_2}{C'_1}, \sigma'
}
{
	C_1, \sigma \ \to \   C'_1, \sigma'
}
\\[1ex]
&\infer[\rname{C}{Struct}]
{
	C_1, \sigma \to  C'_1, \sigma'
}
{
	C_1 \precongr C_2
	& C_2, \sigma \to   C'_2, \sigma'
	& C'_2  \precongr C'_1
}
\end{eqnarray*}
\caption{Core Choreographies, Semantics.}
\label{fig:cc_semantics}
\end{figure}
These rules formalise the intuition presented earlier.
In the premise of $\rname{C}{Com}$, we write $e[\sigma(\pid p)/\pcont]$ for the result of replacing $\pcont$ with $\sigma(\pid p)$ in $e$.
In the reductum, $\sigma[\pid q \mapsto v]$ denotes the updated state function $\sigma$ where $\pid q$ now maps to $v$.

Rule \rname{C}{Struct} uses the structural precongruence relation $\precongr$, which gives a concurrent 
interpretation to choreographies by allowing non-interfering actions to be executed in any order.
The key rule defining $\precongr$ is
\[
\infer[\rname{C}{Eta-Eta}]
{\eta;\eta'\ \equiv \ \eta';\eta}
{\pn(\eta) \cap \pn(\eta') = \emptyset}
\]
where $C \equiv C'$ stands for $C \precongr C'$ and $C' \precongr C$ and $\pn(C)$ returns the set of all process names occurring in $C$.
The other rules for $\precongr$ are standard, and support recursion unfolding and garbage collection of
unused definitions.

CC was designed as a core choreography language, in which in particular it is possible to implement any computable 
function. Furthermore, CC choreographies can always progress until they terminate.

\begin{theorem}
\label{thm:df-by-design}
If $C$ is a choreography, then either
$C \precongr \nil$ ($C$ has terminated)
or, for all $\sigma$, $C,\sigma \to C',\sigma'$ for some $C'$ and $\sigma'$ ($C$ can reduce).
\end{theorem}

Label selections are not required for Turing completeness, and thus the simpler fragment MC 
obtained from CC by omitting them is interesting as an intermediate language for compilers and, also, for theoretical 
analysis.
One of the reasons for having label selection is to make choice propagation explicit in choreographies; in a system 
implementation, this allows, e.g., to monitor distributed choices without having to inspect the message payload.
Another reason is \emph{projectability}: the possibility of automatically generating 
processes implementations that satisfy the choreographic specification.
In Example~\ref{ex:abs}, the label selections in Lines 5 and 7 are important in order for $\pid b$ to let $\pid a$ and 
$\pid s$ know whether or not they should communicate.

Choices communicated by label selections can also be encoded as data in value communications, by sending a 
boolean value to determine which one of two branches was selected.
This is the key idea behind the encoding presented in~\cite{ourstuff} -- arrow ($a$) in 
Figure~\ref{fig:ouroboros} -- which transforms a choreography in CC to one in MC by encoding selections as value 
communications and nested conditionals.

We do not need to concern ourselves with projectability in this work, and we will thus omit its details.
This is because CC and MC enjoy a projectability property that is not altered by our development.
Formally, there exists a procedure $\amend(\cdot)$ that, given any choreography, returns a choreography in CC that is 
projectable. Then, given a projectable CC choreography, the encoding $\delsel\cdot$ transforms it into a choreography 
in MC, by encoding selections as value communications and conditionals.
These transformations preserve the computational meaning of choreographies, as formally stated in the following theorem 
($\cdot^+$ extends a state function to the auxiliary processes introduced by the transformations in a systematic way).
\begin{theorem}
  \label{thm:delsel}
  Let $C,C'$ be MC choreographies and $\sigma,\sigma'$ be states.
  If $C,\sigma\to^\ast C',\sigma'$, then $\delsel{\amend(C)}^+,\sigma^+\to^\ast\delsel{\amend(C')},{\sigma'}^+$.
\end{theorem}

The main limitation of CC is that its semantics is synchronous.
Indeed, in a real-world scenario implementation of Example~\ref{ex:abs}, we would expect $\pid s$ to proceed immediately to sending its message in Line 3 after having sent the one in Line 2, without waiting for $\pid a$ to receive the latter.
Capturing this kind of asynchronous behaviour is the main objective of our development in the remainder of this paper.


\section{Asynchrony in MC}
\label{sec:asynchrony}

In this section, we extend CC with primitives to implement asynchronous communication, obtaining a calculus of Dynamic Core Choreographies (DCC).
We focus on MC and first show that any MC choreography can be encoded in DMC -- the fragment of DCC that does not use label selection -- in such a way that communication becomes asynchronous.

More precisely, we provide a mapping $\async\cdot{}:\mbox{MC}\to\mbox{DMC}$ such that every communication
action $\gencom\in C\in\mbox{MC}$ becomes split into a send/receive pair in $\async C{}\in\mbox{DMC}$, with the
properties that: $\pid p$ can continue executing without waiting for $\pid q$ to receive its message (and even
send further messages to $\pid q$); and messages from $\pid p$ to $\pid q$ are delivered in the same order as
they were originally sent.

\smallpar{The system DCC.}
We briefly motivate DCC.
In CC, there is a bound on the number of values that can be stored at any given time by the
system: since each process can hold a single value, the maximum number of values the system can know is equal
to the number of processes in the choreography, which is fixed.
However, in an asynchronous setting, the number of values that need to be stored is unbounded:
a process $\pid p$ may loop forever sending values to $\pid q$, and $\pid q$ may wait an arbitrary long time
before receiving any of them.
Therefore, we need to extend CC with the capability to generate new processes.
As discussed in~\cite{ourPCstuff}, this requires enriching the language with two additional abilities: parameters to
recursive procedures (in order to be able to use a potentially unbounded number of processes at the same time)
and action to communicate process names.

Formally, the differences between the syntax of CC and that of DCC are highlighted in Figure~\ref{fig:dcc_syntax}:
procedure definitions and calls now have parameters; there is a new term for generating processes; and, the 
expressions sent by processes can also be process names.
\begin{figure}
\begin{align*}
C ::={} & \cdots 
  \acspar \genrecp \acspar \gencallp
\\
\eta ::={} & \cdots 
  \acspar \genstart
\qquad
e ::= \pid p \acspar
  \cdots 
\end{align*}
\caption{Dynamic Core Choreographies, Syntax.}
\label{fig:dcc_syntax}
\end{figure}
The possibility of communicating a process name ($\com{\pid p.\pid q}{\pid r}$) ensures name mobility.
We will use the abbreviation $\gentell$ as shorthand for
$\com{\pid p.\pid q}{\pid r};\,\com{\pid p.\pid r}{\pid q}$.

The semantics for DCC includes an additional ingredient, borrowed from~\cite{ourPCstuff}: a graph of connections $G$, 
keeping track of which pairs of processes are allowed to communicate.
This graph is directed, and an edge from $\pid p$ to $\pid q$ in $G$ (written $\know{\pid p}{\pid q}$) means that $\pid p$ knows the name of $\pid q$.
In order for an actual message to flow between $\pid p$ and $\pid q$, both processes need to know each other, which we 
write as $\knows{\pid p}{\pid q}$.\footnote{%
In some process calculi, the weaker condition $\know{\pid p}{\pid q}$ is typically sufficient for $\pid p$ to send a 
message to $\pid q$. Our condition is equivalent to that found in the standard model of Multiparty Session 
Types~\cite{HYC16}. This choice is orthogonal to our development.
}
The reduction relation now has the form $G,C,\sigma\to G',C',\sigma'$, where $G$ and $G'$ are the connection graphs before and after executing $C$, respectively.
The complete rules are given in Figure~\ref{fig:dcc_semantics}, with $\precongr$ defined similarly to CC. In 
rule~\rname{C}{Start}, the fresh process $\pid q$ is assigned a default value $\bot$.

\begin{figure}
\begin{eqnarray*}
&\infer[\rname{C}{Com}]
{
  G,\gencom;C,\sigma
  \ \to \ 
  G, C, \sigma[\pid q \mapsto v]
}
{
  \knows{\pid p}{\pid q} &
  \eval{e[\sigma(\pid p)/\pcont]}v
}
\\[1ex]
&\infer[\rname{C}{Sel}]
{
  G,\gensel;C,\sigma \ \to \ G,C,\sigma
}
{
  \knows{\pid p}{\pid q}
}
\\[1ex]
&\infer[\rname{C}{Start}]{
  G, \genstart;C, \sigma
  \ \to\ 
  \update{\pid p}{\pid q},C, \sigma[\pid q\mapsto\bot]
}{}
\\[1ex]
&\infer[\rname{C}{Intro}]{
  G, \genintro;C, \sigma
  \ \to\ 
  \updates{\pid q}{\pid r}, C, \sigma
}{
  \knows{\pid p}{\pid q}
  &
  \know{\pid p}{\pid r}
}
\\[1ex]
&\infer[\rname{C}{Cond}]
{
  G, \gencond, \sigma \ \to \ G, C_i, \sigma
}
{
  i = 1 \ \text{if } \sigma(\pid p) = \sigma(\pid q), &
  i = 2 \ \text{otherwise}
}
\\[1ex]
&\infer[\rname{C}{Struct}]
{
  G, C_1, \sigma \ \to \  G', C'_1, \sigma'
}
{
  C_1\, \precongr \,C_2
  & G, C_2, \sigma\ \to \  G', C'_2, \sigma'
  & C'_2 \, \precongr\, C'_1
}
\end{eqnarray*}
\caption{Dynamic Core Choreographies, Semantics.}
\label{fig:dcc_semantics}
\end{figure}

The proof of Theorem~\ref{thm:df-by-design} can be generalised to DCC, but this requires an extra ingredient: a simple 
type system (which we do not detail, as it is a subsystem of that presented in~\cite{ourPCstuff}). This type system 
checks that all processes that attempt at communicating are connected in the communication graph, e.g., by being 
properly introduced using name mobility.
Furthermore, we can define a target process calculus for DCC and an EndPoint Projection that will automatically synthetise correct-by-construction deadlock-free implementations of (projectable) choreographies, using techniques from~\cite{ourstuff} and~\cite{ourPCstuff}.
Although these constructions are not technically challenging, we omit them for brevity, since they are immaterial for our results.

The fragment of DCC that does not contain label selections is called Dynamic Minimal Choreographies (DMC).
Amendment and label selection elimination hold for DMC and DCC just as for MC and CC, so that, for any DMC choreography $C$, $\delsel{\amend(C)}$ is always projectable, and Theorem~\ref{thm:delsel} also holds for these calculi (arrow ($b$) in Figure~\ref{fig:ouroboros}).

\smallpar{The encoding.}
We focus now on the mapping ($1$) from Figure~\ref{fig:ouroboros}, as this is the key ingredient to establish the remaining connections in that figure.
Let $C$ be a choreography in MC.
In order to encode $C$ in DMC, we use a function $M_C:\mathcal P^2\to\NN$, where $\mathcal P=\pn(C)$ is
the set of process names in $C$.
Intuitively, $\async C{}$ will use a countable set of auxiliary processes
$\left\{\pid{pq}^i\mid\pid p,\pid q\in\mathcal P,i\in\NN\right\}$, where $\pid{pq}^i$ will hold the $i$th
message from $\pid p$ to $\pid q$.

First, we setup initial channels for communications between all processes occurring in $C$.
\[
  \async C{} =
  \left\{\pid p\,\m{start}\,\pid{pq}^0;\,\pid p:\pid q\,\code{<->}\pid{pq}^0\right\}_{p,q\in\mathcal P,\pid p\neq\pid 
q};\,
  \async C{M_0}\]
Here, $M_0(\pid p,\pid q)=0$ for all $\pid p$ and $\pid q$.
For simplicity, we write $\pid{pq}^M$ for $\pid{pq}^{M(\pid p,\pid q)}$ and $\pid{pq}^{M+}$ for
$\pid{pq}^{M(\pid p,\pid q)+1}$.
The definition of $\async CM$ is given in Figure~\ref{fig:async}.

\begin{figure*}
\begin{align*}
    \async{\gencom;C}M ={} & \com{\pid p.e}{\pid{pq}^M};\,
                        \start{\pid p}{\pid{pq}^{M+}};\,
                        \pid p:\pid{pq}^M\,\code{<->}\,\pid{pq}^{M+};\,\\
                        &
			\pid{pq}^M.\pid q \code{->} \pid{pq}^{M+};\,
			\pid{pq}^M.\pid{pq}^{M+} \code{->} \pid{q};\,
			\com{\pid{pq}^M.\pcont}{\pid q};\,\\
                        &
                        \async{C}{M[(\pid p,\pid q)\mapsto M(\pid p,\pid q)+1]} \\
  \async{\gencond}M ={} & \com{\pid q.\pcont}{\pid{qp}^M};\,
                        \start{\pid q}{\pid{qp}^{M+}};\,
                        \pid q:\pid{qp}^M\,\code{<->}\,\pid{qp}^{M+};\,\\
                        &\pid{qp}^M.\pid p\,\code{->}\,\pid{qp}^{M+};\,
                        \pid{qp}^M.\pid{qp}^{M+}\,\code{->}\,\pid p;\\
                        &\m{if}\,\eqcom{\pid p}{\pid{qp}^M}\,\m{then}\,\async{C_1}{M[(\pid p,\pid q)\mapsto M(\pid p,\pid q)+1]}\,\\
                        &\hspace*{4.3em}\m{else}\,\async{C_2}{M[(\pid p,\pid q)\mapsto M(\pid p,\pid q)+1]}\\
  \async\nil M ={} &\nil \\
  \async{X}M ={} & X\langle\bar M\rangle \\
  \async{\genrec}M ={} & \m{def}\,X(\overline{M_0})=\async{C_2}{M_0}\,\m{in}\,\async{C_1}M
\end{align*}
\caption{Encoding MC in DMC}
\label{fig:async}
\end{figure*}
We write $\bar M$ for $\left\{\pid{pq}^M\mid\pid p,\pid q\in\mathcal P,\pid p\neq\pid q\right\}$, where we
assume that the order of the values of $M$ is fixed.
In recursive definitions, we reset $M$ to $M_0$; note that the parameter declarations act as binders, so these
process names are still fresh.
We can use $\alpha$-renaming on $\async{C}{}$ to make all bound names distinct.

In order to encode $\gencom$, $\pid p$ uses the auxiliary process $\pid{pq}^M$ to store the value it wants to send to $\pid q$.
Then, $\pid p$ creates a fresh process (to use in the next communication) and sends its name to $\pid{pq}^M$.
Afterwards, $\pid p$ is free to proceed with execution.
In turn, $\pid{pq}^M$ communicates $\pid q$'s name to the new process, which now is ready to receive the next message from $\pid p$.
Finally, $\pid{pq}^M$ waits for $\pid q$ to be ready to receive both the value being communicated and the name of the process that will store the next value.

The behaviours of the choreographies $C$ and $\async C{}$ are closely related, as formalised in the following theorems.
\begin{theorem}
  \label{lem:async1}
  Let $\pid p\in\pn(C)$ and $\pid{pq}\in\pn(\async C{})\setminus\pn(C)$.
  If $G,\async{C}{},\sigma\to^\ast G',C_1,\sigma_1\to G',C_2,\sigma_2$ where in the last transition a value $v$ is
  sent from $\pid p$ to $\pid{pq}$, then there exist $G'',C_3,\sigma_3,C_4$ and $\sigma_4$ such that
  $G',C_2,\sigma_2\to^\ast G'',C_3,\sigma_3\to G'',C_4,\sigma_4$ and in the last transition the same value $v$ is sent
  from $\pid{pq}$ to some process $\pid q\in\pn(C)$.
\end{theorem}

Theorem~\ref{lem:async1} states that messages sent from $\pid p$ to $\pid q$ are eventually received by 
$\pid q$.


\begin{theorem}
  \label{lem:async2}
  If $G,\async C{M},\sigma\to^\ast G_1,C_1,\sigma_1$, then there exist $G''$, $C'$, $\sigma'$ and $\sigma''$ such that
  $C,\sigma\to^\ast C',\sigma'$, and $G_1,C_1,\sigma_1\to^\ast G'',\async{C'}{M},\sigma''$, and $\sigma'$ and $\sigma''$
  coincide on the values stored at $\pn(C)$.
\end{theorem}

Theorem~\ref{lem:async2} states that the encoding does not add any additional behaviour to the original choreography, 
aside from expanding communications into several actions.

\begin{example}
  \label{ex:abs_enc}
  We partially show the result of applying this transformation to Lines 1--3 of Example~\ref{ex:abs}.
  We only include the initializations of the channels that are used in this fragment; the numbers indicated refer to the line numbers in the original example.
  \begin{align*}
    & \start{\pid a}{\pid{as}^0};\,\tell{\pid a}{\pid s}{\pid{as}^0}; \\
    & \start{\pid s}{\pid{sa}^0};\,\tell{\pid s}{\pid a}{\pid{sa}^0}; \\
    & \start{\pid s}{\pid{sb}^0};\,\tell{\pid s}{\pid b}{\pid{sb}^0}; \\
    1.\quad
    & \com{\pid a.\mathit{title}}{\pid{as}^0};\,
      \start{\pid a}{\pid{as}^1};\,
      \tell{\pid a}{\pid{as}^0}{\pid{as}^1};\,\\
    & \intro{\pid{as}^0}{\pid s}{\pid{as}^1};\,
      \intro{\pid{as}^0}{\pid{as}^1}{\pid s};\,
      \com{\pid{as}^0.\pcont}{\pid s};\\
    2.\quad
    & \com{\pid s.\mathit{price}}{\pid{sa}^0};\,
      \start{\pid s}{\pid{sa}^1};\,
      \tell{\pid s}{\pid{sa}^0}{\pid{sa}^1};\,\\
    & \intro{\pid{sa}^0}{\pid a}{\pid{sa}^1};\,
      \intro{\pid{sa}^0}{\pid{sa}^1}{\pid a};\,
      \com{\pid{sa}^0.\pcont}{\pid a};\\
    3.\quad
    & \com{\pid s.\mathit{price}}{\pid{sb}^0};\,
      \start{\pid s}{\pid{sb}^1};\,
      \tell{\pid s}{\pid{sb}^0}{\pid{sb}^1};\,\\
    & \intro{\pid{sb}^0}{\pid b}{\pid{sb}^1};\,
      \intro{\pid{sb}^0}{\pid{sb}^1}{\pid b};\,
      \com{\pid{sb}^0.\pcont}{\pid b};\\
    & \ldots
  \end{align*}
  The first three lines initialize three channels: from $\pid a$ to $\pid s$; from $\pid s$ to $\pid a$; and from $\pid s$ to $\pid b$.
  Then one message is passed in each of these channels, as dictated by the encoding.
  All communications are asynchronous in the sense explained above, as in each case the main sender process sends its message to a dedicated intermediary ($\pid{as}^0$, $\pid{sa}^0$ or $\pid{sb}^0$, respectively), who will eventually deliver it to the recipient.
  Moreover, causal dependencies are kept: in Step~2, $\pid s$ can only send its message to $\pid{sa}^0$ after receiving the message sent by $\pid a$ in Step~1.
  However, in Step~3 $\pid s$ can send its message to $\pid{sb}^0$ without waiting for $\pid a$ to receive the previous 
message, as the action $\com{\pid s.\mathit{price}}{\pid{sa}^0}$ can swap with the three actions immediately preceding it.

  We briefly illustrate Theorems~\ref{lem:async1} and~\ref{lem:async2} in this setting.
  Theorem~\ref{lem:async1} states that, e.g., the action $\com{\pid a.\mathit{title}}{\pid{as}^0}$ is eventually 
followed by a 
communication of $\mathit{title}$ from $\pid{as}^0$ to some other process in the original choreography (in this case, $\pid s$).
  Theorem~\ref{lem:async2} implies that if, e.g., $\com{\pid a.\mathit{title}}{\pid{as}^0}$ is executed, then it must 
be ``part'' 
of an action in the original choreography (in this case, $\com{\pid a.\mathit{title}}{\pid s}$), and furthermore it is possible 
to find an execution path that will execute the remaining actions generated from that one (the remaining 
five actions in Step~1).
\end{example}
%


\section{The General Case}
\label{sec:general}

The calculus DMC is in itself synchronous, just like MC.
We now show that we can extend $\async\cdot{}$ to the full language of DMC -- arrow ($2$) in 
Figure~\ref{fig:ouroboros} --
thereby obtaining a systematic way to write asynchronous communications in DMC.
By further marking which communications we want to treat as synchronous (so that they are untouched by
$\async\cdot{}$) we obtain a calculus in which we can have both synchronous and asynchronous communication,
compiled in itself.
This is similar (albeit dual) to the situation in asynchronous $\pi$-calculus, where we can also encode
synchronous communication without extending the language.

The main challenge is dealing with $M$, as the source choreography can now include process spawning.
This means that the domain of $M$ can be dynamically extended throughout the computation of $\async CM$, which
renders our parameter-passing in recursive calls invalid (since the number of parameters in the procedures generated 
by our encoding is fixed).
However, since each procedure $X(\pid p_1,\ldots,\pid p_n)$ in DMC can only use (by convention) the processes 
$\pid p_1,\ldots,\pid p_n$ in its body, we can restrict the additional
parameters introduced by the encoding to the $n(n-1)$ auxiliary processes currently assigned by $M$ to communications 
between the $\pid p_i$s. For example,
$\async{\rec{X(\pid p,\pid q)}{C_2}{C_1}}{M}$ would be
$\rec{X(\pid p, \pid q, \pid{pq}^0, \pid{qp}^0)}{\async{C_2}{M_0}}{\async{C_1}{M}}$.
We will not write this definition formally.

With this in mind, we can easily define the new cases for $\async CM$.
\begin{align*}
  \async{\start{\pid p}{\pid q};C} M ={} & \start{\pid p}{\pid q};\,
                                           \start{\pid p}{\pid{pq}^0};\,
                                           \start{\pid q}{\pid{qp}^0};\\
                                           &\pid p:\pid q\,\code{<->}\,\pid{pq}^0;\,
                                           \pid q:\pid p\,\code{<->}\,\pid{qp}^0;\\
                                           &\async{C}{M[(\pid p,\pid q)\mapsto 0,(\pid q,\pid p)\mapsto 0]} \\
  \async{\pid p.\pid q\,\code{->}\,\pid r;C}M ={} &
                        \start{\pid p}{\pid{qr}^0};\,
                        \pid p.\pid{qr}^0\,\code{->}\,\pid{pq}^M;\,
                        \pid p.\pid{qr}^0\,\code{->}\,\pid{pr}^M;\\
                        &
                        \pid p\,\m{start}\,\pid{pq}^{M+};\,
                        \pid p:\pid{pq}^M\,\code{<->}\,\pid{pq}^{M+};\\
                        &
                        \pid p\,\m{start}\,\pid{pr}^{M+};\,
                        \pid p:\pid{pr}^M\,\code{<->}\,\pid{pr}^{M+};\\
                        &
                        \pid{pq}^M.\pid q\,\code{->}\,\pid{pq}^{M+};\,
                        \pid{pr}^M.\pid r\,\code{->}\,\pid{pr}^{M+};\\
                        &
                        \pid{pq}^M.\pid{pq}^{M+}\,\code{->}\,\pid{q};\,
                        \pid{pq}^M.\pid{qr}^0\,\code{->}\,\pid q;\\
                        &
                        \pid{pr}^M.\pid{pr}^{M+}\,\code{->}\,\pid r;\,
                        \pid{pr}^M.\pid{qr}^0\,\code{->}\,\pid r;\\
                        &\async{C}{M[(\pid p,\pid q)\mapsto M(\pid p,\pid q)+1,(\pid p,\pid r)\mapsto M(\pid p,\pid r)+1,(\pid q,\pid r)\mapsto 0]}
\end{align*}
In $\async{\start{\pid p}{\pid q};C}M$, we simply create the asynchronous communication channels between
$\pid p$ and $\pid q$ -- the only step where these process will need to synchronize -- and extend $M$ in the
continuation.
The encoding of $\pid p.\pid q\,\code{->}\,\pid r$ is better understood by reading it as a composition:
first, $\pid p$ creates the new asynchronous communication channel from $\pid q$ to $\pid r$, then
uses its own channels to send this name to these processes.
Note that the auxiliary channels do not communicate, so this encoding will introduce asymmetries in the graph
of communications.

Theorems~\ref{lem:async1} and~\ref{lem:async2} still hold for this extended encoding.

\smallpar{Projections.}
Finally, we extend this encoding to the whole language of DCC -- arrow ($4$) in Figure~\ref{fig:ouroboros} -- by adding 
the clause
\begin{align*}
  \async{\gensel;C}M ={} &
                         \com{\pid p}{\pid{pq}^M[l]};\,
                         \start{\pid p}{\pid{pq}^{M+}};\\
                         &
                         \pid p:\pid{pq}^M\,\code{<->}\,\pid{pq}^{M+};\,
                         \pid{pq}^M.\pid q\,\code{->}\,\pid{pq}^{M+};\\
                         &
                         \pid{pq}^M.\pid{pq}^{M+}\,\code{->}\,\pid q;\,
                         \com{\pid{pq}^M}{\pid q[l]};\\
                         &
                         \async{C}{M[(\pid p,\pid q)\mapsto M(\pid p,\pid q)+1]}
\end{align*}
to the definition of $\async CM$.
Restricting this encoding to the language of CC yields arrow ($3$) in Figure~\ref{fig:ouroboros}.

We finish this section with a brief informal note on projectability.
As we discussed in \S~\ref{sec:mc}, a formal presentation of projection for DMC and DCC is beyond the scope of this 
paper. However, we point out that our encoding for asynchronous communications preserves projectability, 
i.e., if $C$ is projectable, then so is $\async C{}$.


\section{Related Work}
\label{sec:related}
%

To the best of our knowledge, this is the first work presenting an interpretation of asynchronous communications in 
choreographies based solely on the expressive power of primitives for the creation of processes and their 
connections, via name mobility.

Our work recalls the development of the asynchronous $\pi$-calculus~\cite{HT91} (A$\pi$ for short, using the 
terminology from~\cite{SW01}). A$\pi$ has a synchronous semantics, in the sense that two processes can communicate when 
they are both ready to, respectively, perform compatible input and output actions. However, an output action can have 
no sequential continuation, but can instead only be composed in parallel with other behaviour. Thus, the interpretation 
of communications in A$\pi$ is asynchronous, since outputs can be seen as messages in transit over a network. The 
synchronisation between (the process holding) a message in transit and the intended receiver models then the 
extraction of the message from the medium by the receiver.
Differently from our work, A$\pi$ is obtained from the standard $\pi$-calculus by \emph{restricting} the syntax of 
processes such that all communications necessarily conform to this asynchronous interpretation.
It is then shown that A$\pi$ is expressive enough to encode the synchronous communications from standard $\pi$-calculus,
by using acknowledgement messages. DMC and DCC exhibit the dual behaviour: communications are naturally synchronous, 
but we can always encode them to be asynchronous by passing them through intermediary processes.

Other studies have investigated asynchronous communications in choreographies. The distinctive feature of 
our work is that it does not rely on any ad-hoc syntax or semantics for capturing asynchrony.
In~\cite{HYC08}, choreographies are used as types for communication protocols and are related to asynchronous 
communications by encoding choreographies in types for terms in a variant of the $\pi$-calculus. However, asynchrony 
can only be observed in the semantics of processes, not at the level of choreographies, and the syntax of processes is 
equipped with ad-hoc runtime terms\footnote{Runtime terms are assumed never to be used by the 
programmer, but only produced as the result of execution.} that represent messages in transit. The first 
work defining an asynchronous semantics for choreographies is~\cite{CM13}, by defining an ad-hoc rule in the semantics 
of choreographies that allows nested (not appearing at the top level) communications to be executed if, among other 
conditions: the sender is the same as the one in the communication at the top level of the choreography, the 
receiver is not involved in the nested communication. This technique has been later adopted also in~\cite{MY13} -- for 
defining the composition of asynchronous choreographies with legacy process code -- and in~\cite{HYC16} (the journal 
version of~\cite{HYC08}) -- to formulate a semantics for communication protocols represented as choreographies.
In~\cite{DY13}, choreographies (not processes, for example as in~\cite{HYC08}) are equipped with runtime terms to 
represent messages in transit.

Process spawning and name mobility are the key additions to DCC and DMC, from CC and MC, that yield the expressive 
power to represent asynchronous communications. Process spawning in choreographies has been studied also 
in the works~\cite{CHY12,CM13,MY13}, but in a different form where processes have to synchronise over a shared channel 
to proceed. Name mobility in choreographies was introduced in~\cite{CM13}, but for channel rather than process names. 
Our process spawning and name mobility primitives are simplifications of those presented in~\cite{ourPCstuff}, which 
makes all results from that work applicable to DCC (and thus DMC).

%
%


\section{Conclusions}
\label{sec:conclusions}

Choreographies are widely used in the context of concurrent and distributed software architectures, in order to specify 
precisely how the different components of a system should interact~\cite{BPMN,WSCDL}.
Previous formalisations of asynchronous communications in choreographies exchange expressivity for simplicity, yielding 
ad-hoc models with unclear connections.
In this work, we showed that a choreography calculus with process spawning and process name mobility can capture 
asynchronous communications. Therefore, all such calculi with similar primitives have the same power. Our 
development is conservative wrt previous work, allowing us to import existing techniques developed for 
previous calculi. For example, the techniques shown in~\cite{ourPCstuff,ourstuff} could be reapplied to DCC to 
synthesise deadlock-free process implementations. Here, we showed how to import the result of selection elimination 
from~\cite{ourstuff}.
In conclusion, we now have a setting where we can reason about asynchronous communications in choreographies by 
considering a simple synchronous semantics, just like it can be done in the seminal model of $\pi$-calculus for mobile 
processes.
This work extends results previously published in~\cite{sac17}.


\paragraph{Acknowledgements.}
Montesi was supported by CRC (Choreographies for Reliable and efficient Communication software),
grant no.\ DFF--4005-00304 from the Danish Council for Independent Research.

\bibliographystyle{abbrv}
\bibliography{biblio}

\end{document}